\documentclass[11pt]{article}  
\usepackage{amsmath}  
\usepackage{pstricks}
\usepackage{hyperref}
 \hypersetup{
     colorlinks=true,
     linkcolor=black,
     filecolor=back,
     citecolor=blue,      
     urlcolor=cyan,
     }
\usepackage[most]{tcolorbox}

\usepackage[vcentermath]{youngtab}
\usepackage{bbm}
\usepackage{amssymb,amsfonts}

\newcommand{\re}[1] {(\ref{#1})}

\def\half{\frac{1}{2}}

\newfont{\bbbold}{msbm10 scaled \magstep1}

\def\cL{{\cal L}}

\def\cN{{\cal N}}
\def\cO{{\cal O}}

\newfont{\goth}{eufm10 scaled \magstep1}

\def\gg{\mbox{\goth g}}

\def\gl{\mbox{\goth l}}

\def\gs{\mbox{\goth s}}

\def\gu{\mbox{\goth u}}

\def\c{\gamma}\def\C{\Gamma}
\def\d{\delta}

\def\i{\iota}

\def\be{\begin{equation}}\def\ee{\end{equation}}
\def\bea{\begin{eqnarray}}\def\eea{\end{eqnarray}}
\def\barr{\begin{array}}\def\earr{\end{array}}

\def\o{\omega}\def\O{\Omega}
\def\del{\partial}

\def\nn{\nonumber}
\def\bd{\begin{document}}
\def\ed{\end{document}}
\def\ba{\begin{array}}
\def\ea{\end{array}}
\def\bea{\begin{eqnarray}}
\def\eea{\end{eqnarray}}
\def\ft#1#2{\tfrac{#1}{#2}}
\def\fft#1#2{\frac{#1}{#2}}
\def\sst#1{{\scriptscriptstyle #1}}
\def\oneone{\rlap 1\mkern4mu{\rm l}}

\newcommand{\eq}[1]{(\ref{#1})}
\newcommand{\w}[1]{\\[0.#1cm]}
\def\eqs#1#2{(\ref{#1}-\ref{#2})}
\def\det{{\rm det\,}}
\def\tr{{\rm tr}}
\def\ad{{\rm ad}}

\newcommand{\hoch}[1]{$\, ^{#1}$}
\newcommand{\imperial}{\it\small Theoretical Physics Group, Imperial College London\\ Prince Consort Road, London SW7 2AZ, UK}
\newcommand{\kings}
{\it\small Department of Mathematics, King's College, University of London\\ Strand, London WC2R 2LS, UK}
\newcommand{\uu}
{\it\small Department of Theoretical Physics, Uppsala, Sweden}
\newcommand{\hip}
{\it\small HIP-Helsinki Institute of Physics, P.O. Box 64 FIN-00014
University of Helsinki, Suomi-Finland}
\newcommand{\stock}
{\it\small Department of Theoretical Physics, Stockholm, Sweden}
\newcommand{\golm}
{\it\small AEI, Max Planck Institut f\"ur Gravitationsphysik\\ Am M\"{u}hlenberg 1, D-14476 Potsdam, Germany}
\makeatletter
\renewcommand\theequation{\thesection.\arabic{equation}}
\@addtoreset{equation}{section} \makeatother

\newcommand{\sa}{/ \hspace{-1.2ex}}
\newcommand{\saa}{/ \hspace{-1.4ex}}
\newcommand{\saaa}{\, / \hspace{-1.6ex}}
\newcommand{\Scal}[1]{\Bigl ({#1} \Bigr )}
\newcommand{\scal}[1]{\bigl ({#1} \bigr )}

\newcommand{\CR}{\nonumber \\*}

\newcommand{\trace}{\hbox {tr}~}
\newcommand{\traceS}{\hbox {tr}_{\scriptscriptstyle \mathfrak{S}}~}

\DeclareMathAlphabet{\mathpzc}{OT1}{pzc}{m}{it}
\def\BRST{\,\mathpzc{s}\,}
\def\aBRST{{\scriptstyle (\mathpzc{s})}}
\def\q{{{\scriptscriptstyle (Q)}}}
\def\qs{{\scriptscriptstyle (Q\mathpzc{s})}}
\def\Qsla{{\mathcal{S}_{\q}}}
\def\Slav{{\mathcal{S}_\aBRST}}
\def\epsilonb{{\overline{\epsilon}}}
\def\bulletup{{\scriptstyle \bullet}}

\newcommand{\gra}[2]{{\scriptscriptstyle (#1 , #2 )}}
\newcommand{\ord}[1]{{\scriptscriptstyle (#1)}}

\newcommand{\na}{\nabla}

\newcommand{\ber}{\begin{eqnarray}}
\newcommand{\eer}[1]{\label{#1}\end{eqnarray}}
\newcommand{\eero}{\end{eqnarray}}

\def\cL{{\cal L}}
\def\cN{\mathcal{N}}
\def\cO{\mathcal{O}}

\def\ie{{\it i.e.}\ }
\def\eg{{\it e.g.}\ }

\newcommand{\sfrac}[2]{{\scriptstyle \frac{#1}{#2}}}
\newcommand{\stfrac}[2]{{\scriptscriptstyle \frac{#1}{#2}}}

 \def\balpha{{\overline{\alpha}}}
 \def\bbeta{{\overline{\beta}}}
 \def\bgamma{{\overline{\gamma}}}
 \def\bdelta{{\overline{\delta}}}
 \def\bepsilon{{\overline{\epsilon}}}
 \def\bvarepsilon{{\overline{\varepsilon}}}
 \def\bzeta{{\overline{\zeta}}}
 \def\bareta{{\overline{\eta}}}
 \def\btheta{{\overline{\theta}}}
 \def\bvartheta{{\overline{\vartheta}}}
 \def\biota{{\overline{\iota}}}
 \def\bkappa{{\overline{\kappa}}}
 \def\blambda{{\overline{\lambda}}}
 \def\bmu{{\overline{\mu}}}
 \def\bnu{{\overline{\nu}}}
 \def\bxi{{\overline{\xi}}}
 \def\bpi{{\overline{\pi}}}
 \def\brho{{\overline{\rho}}}
 \def\bvarrho{{\overline{\varrho}}}
 \def\bsigma{{\overline{\sigma}}}
 \def\bvarsigma{{\overline{\varsigma}}}
 \def\btau{{\overline{\tau}}}
 \def\bphi{{\overline{\phi}}}
 \def\bvarphi{{\overline{\varphi}}}
 \def\bchi{{\overline{\chi}}}
 \def\bpsi{{\overline{\psi}}}
 \def\bomega{{\overline{\omega}}}

\def\thalf{{\textrm{\tiny\textonehalf}}}
\def\tquarter{{\textrm{\tiny\textonequarter}}}
\def\Ko{{\scriptscriptstyle K}}
\def\tKo{\scriptscriptstyle k }
\def\corr{$\clubsuit$}

\newcommand{\auth}{\large P.S.\ Howe${}^{a,}$\footnote{email: paul.howe@kcl.ac.uk} and U. Lindstr\"om${}^{b,}$\footnote{email: ulf.lindstrom@physics.uu.se}}

\begin{document}

\renewcommand{\thefootnote}{\fnsymbol{footnote}}

\null
\begin{flushright}
{\small UUITP-XX/24}
\vskip 1.5 cm
\end{flushright}

\begin{center}
{\Large{\bf Exotic gravity theory in loop space}}
\vspace{.75cm}

\auth
\end{center}
\vspace{.5cm}

\centerline{${}^a${\it \small Department of Mathematics, King's College London}}
\centerline{{\it \small The Strand, London WC2R 2LS, UK}}
\vspace{.5cm}

\centerline{${}^b${\it \small Department of Physics and Astronomy, Theoretical Physics, Uppsala University}}
\centerline{{\it \small SE-751 20 Uppsala, Sweden }}
\centerline{\it \small and}
\centerline{\it \small Centre for Geometry and Physics, Uppsala University,
SE-75106 Uppsala, Sweden
}

\vspace{1cm}


\centerline{{\bf Abstract}}
\vskip .5cm
An exotic linearised theory of superconformal gravity in $D=6, (4,0)$ superspace, proposed by C. Hull, is discussed in loop space with focus on its bosonic sector. Pursuing an analogy to the loop space version  the system of a two form $B$ and its field strength $H$, we show that the exotic linearised gravitational potential $C$ and curvature $G$ can be reinterpreted as an ultra-local (linearised) metric and curvature on loop space. 
\vspace{1cm}

\hfill{\it Dedicated to the memory of Stanley Deser.}

\renewcommand{\thefootnote}{\arabic{footnote}}
\setcounter{footnote}{0}

\pagebreak
\tableofcontents
\setcounter{page}{1}    


\section{Introduction}  

An exotic theory of linearised superconformal gravity has been proposed by C. Hull \cite{Hull:2000rr, Hull:2000zn}. He discusses it for the maximal case of $D=6, (4,0)$ supersymmetry, although the model exists for any $(p,0)$ with $p\leq 4$ in $D=6$.  The leading bosonic component of the potential multiplet is a field $C_{mn,pq}$ in the $(2, 2)$ Young tableau representation of $\gs\gl(6)$ (i.e. 2 boxes in the first row and 2 in the second), where each pair of indices is taken  to be anti-symmetrised.  
The corresponding multiplet reduces to a standard supergravity multiplet in lower dimensions after dualisation. The theory can be thought of as a square of the $(H,B)$-model, where $H=dB$ is a three-form, which exists for $(p,0)$, $p=1,2$, and which was introduced in \cite{Howe:1983fr}  and \cite{Koller:1982cs}. However, not all Hull potentials will be expressible as products of two $B$ fields in the $(2,2)$ Young tableau representation.  A superspace description of Hull's model has been given by Cederwall \cite{Cederwall:2020dui}, while an action has been given for the theory in a $5+1$ split formulation \cite{Bertrand:2020nob, Bertrand:2022pyi}. The field strength $G_{mnp,qrs}$ is obtained from $C$ by acting with 2 derivatives and is in the $(3,3)$ Young tableau representation. This field is self-dual on both sets of indices, a fact which follows from its presentation as the square of $H$. 

In this paper we shall reformulate  the bosonic part of Hull's model in loop space, using similar ideas to those that were employed to discuss $D=10$ heterotic supergeometry in \cite{Bergshoeff:1990mr}, \cite{Howe:1991mf} and \cite{Howe:1991bx},  using the loop space  formalism of \cite{Coquereaux:1988es} . In  {\color{red}\cite{Bergshoeff:1990mr}} it was shown that the 2-form $B$-field in the supergravity multiplet  also gives rise to a 1-form potential in loop superspace.  For the model under discussion, Hull's $C$-field is reinterpreted as a
 two-index tensor in loop space and
 can be interpreted as a metric tensor on loop space, although not in the obvious  sense of lifting a metric on $M$ to one on $LM$. For other types of loop space metrics, see, for example \cite{Maeda:2014koa}.

In the rest of this note we start off with a reformulation of the $D=6$ $(H,B)$ model in loop space, which has not been given before, although it is formally related to the $D=10$ case \cite{Bergshoeff:1990mr}, and then  move on to the exotic supergravity model. 
We shall consider the leading bosonic term, and we shall not need to to be specific about the dimension of spacetime. We shall begin with the $(H,B)$ model, then move on to the $(G,C)$ model.

\section{Loop space}

A loop is a smooth map $\c:S^1\rightarrow M$ and the space of such maps is called loop space, denoted $LM$. The tangent space to a loop $\c$ is $T_{\c}(LM)=\C(\c^*(TM)$. We shall give a very informal discussion of the differential geometry of such spaces in local coordinates following reference \cite{Coquereaux:1988es}. We let $(s,t)$ etc denote coordinates on $S^1$ while local coordinates on $M$ are denoted by $x^m$. So a loop is given by a function $x^m(s)$. We shall be dealing mainly with local tensors on $LM$, i.e fields that are derived from tensors on $M$ in the natural way. For example, if $\o$ is a one-form on $M$ then the corresponding loop form is

\be
\o=\int\, ds\, \d x^m(s) \o_m(x(s))\ ,
\ee
where $\d x^m(s)$ denotes the local basis one-forms on $LM$. Alternatively, one can consider such forms without the circle integration, but we choose to keep it as it makes the parallels between tensors on finite-dimensional spaces and loops spaces manifest. The components of $\o$ are simply $\o_m(x(s))$, but those for higher-rank tensors will also involve delta-functions. For example, the components of a local two-form $\o$ are

\be
\o_{mn}(s,t)[x]=\d(s-t) \o_{mn}(x(s))\ .
\ee

A key r\^{o}le in the formalism is played by the vector field that generates transformations of the circle  which we shall denote by $\dot{X}$, explicitly

\be 
\dot{X}=\int\,ds\, \dot{x}^{m}(s) \frac{\d}{\d x^{m}(s)}\ \ .
\ee
The local forms, and more generally local tensors, are annihilated by the Lie derivative with respect to $\dot{X}$:
\be
{\cal{L}}_{\dot{X}} \o[x(s)]=0 .
\ee

\section{$(B,H)$ model}

.

We can now reformulate this theory in loop space. 
We start from a two-form $B$ on $M$, and let $LM$ denote the corresponding loop space, with circle coordinate $s$. On $LM$  we denote the vector field which generates circle transformations by $\dot X$. We can regard any form on $M$ as an ultra-local form on $LM$, i.e. one which is invariant under transformations generated by $\dot X=\int\,ds\, \dot x^m(s)\frac{\d}{\\d x^m(s)}$. For the $(H,B)$ system on $LM$ we can convert any $p$-form to a $(p-1)$-form by contracting it with  $\dot X$, e.g 
\be
B\rightarrow \tilde B:=i_{\dot X} B\ ,
\ee
from which, for the field strength, we find

\be
\tilde H=d\tilde B =d i_{\dot X} B=-i_{\dot X} d B+{\cal{L}}_{\dot X}B =-i_{\dot X} H\ ,
\ee
where the last step follows from the fact that ultra-local forms on $LM$ are invariant under circle transformations, i.e. annihilated by the Lie derivative ${\cal{L}}_{\dot X}$. The gauge transformation of $\tilde B$ is $d \tilde f$ where $\tilde f=i_{\dot X} f$, $f$ being a one-form. 

The components of a two-form $B$ on loop space are $B_{mn}(s,t)[x]$, 
and in the case of such a form derived from a two-form on $M$ we have

\be
B_{mn}(s,t)[x]=\d(s-t) B_{mn}(x(s))\ ,
\ee
so that

\begin{align}
B=&\frac{1}{2}\int ds\,dt\, \d x^n(t)\wedge\d x^m(s) B_{mn}(s,t)[x]  \nn\w1
 =&\frac{1}{2}\int ds\,\d x^n(s)\wedge\d x^m(s)  B_{mn}(x(s)) \ .
\end{align}
For the induced one-form $\tilde B$ we have, 

\be\label{tilB}
\tilde B=\int ds\  \d x^{m}(s) \dot x^n (s)\, B_{nm}(x(s))\ ,
\ee
while the two-form field-strength is

\be
\tilde H=\half\int ds\ \d x^n(s)\wedge\d x^m(s)~\!\dot x^p H_{pmn} \ .
\ee
\\
As shown in \cite{Howe:1991bx}, this system may be extended to include  Yang-Mills fields. This entails introducing a semisimple gauge group $G$ with algebra generators $T_i$ satisfying
\be
[T_i, T_j ] =c_{ij}^{~~k}T_k
\ee
The Lie-algebra valued gauge field is
\be
A=A^iT_i
\ee
and the set of descent equations are
\bea\nn\label{tfs}
&&H=dB+Q_3~,\\[1mm]\nn
&&P_4=dQ_3~,\\[1mm]\nn
&&\d Q_3=dQ_2(k,A)~,\\[1mm]   
&&\d B=dm+Q_2~,
\eea
where the Chern Simons form is 
\be
Q_3(A)=<A,dA-{\textstyle \frac 2 3} A^2>~,
\ee
where he bracket represents a contraction with the Killing metric,
$P_4$ is a gauge invariant function of the field strength $F$, $m$ and $k=k^iT_i$ are a gauge parameters for the $B$ and $A$ field transformations respectively, and the field strength $H$ is invariant under the transformations in \re{tfs}. 
In loop space the gauge field $A(x)\rightarrow A[x(s)]$
becomes the gauge field for the loop group $LG$. The relation for $H$ in \re{tfs} is still valid, but with
\be
Q_3(A(s))=\int ds<A(s),dA(s)-{\textstyle \frac 2 3} A^2(s)>~.
\ee
Contracting the resulting expression for $H$ with $\dot X$ we have
\be
\tilde H=d\hat B-\int ds [<\i_{\dot X}A(s),F(s)>+{\textstyle \frac 1 2}\!<A(s),A'(s)>]
\ee
where the field strength is $F(s)=dA(s)-A^2(s)$, and 
\be
\hat B=\tilde B+\int  ds <A(s),\i_{\dot X}A(s)>~,
\ee
with $\tilde B$ defined in \re{tilB}.
The gauge transformation of $\hat B$ is
\be
\d \hat B=d\hat m-\int ds <A(s),k'(s)>
\ee
where
\be
\hat m=\tilde m -\int ds <A(s),k(s)>~,
\ee
It is then argued in  \cite{Howe:1991bx} that its transformation properties allow for an interpretation of $\hat B$ as the $U(1)$ part of an $\hat L G$ gauge field on loop space, where 
$\hat L G$ extends the loop group $L G$ by $U(1)$. The Lie algebra of $\hat L G$ is $\hat L \gg=L\gg\oplus \mathbb{R}$. Denoting elements of $\hat L \gg$ by $\hat k=(k(s),b)$ the Lie algebra is
\bea\nn
&&[\hat k_1,\hat k_2]_{L \gg}=[k_1,k_2]_{L \gg}\\[1mm]\nn
&&[\hat k_1,\hat k_2]_{\gu(1)}=[k_1,k_2]_{\gu(1)}\\[1mm]
&&=\int ds <k_1(s),k_2'(s)>
\eea
Similarly, the gauge field for $\hat L G$ is $\hat A=( A(s),\hat B)$. The curvature is
\bea\nn
&&\hat F|_{L\gg}=F\\[1mm]\nn
&&\hat F|_{\gu(1)}=d\hat B-\int  ds <A(s),A'(s)>\\[1mm]
&&=\tilde H+\int  ds <\i_{\dot X}A(s),F(s)>
\eea

\section{$(C,G)$ model}

In order to discuss Hull's $C$-field it will be useful to consider generalised differential forms which correspond to 2-column Young  tableaux with $(p,q)$ columns in the first and second rows respectively \cite{deMedeiros:2002qpr,Howe:2018lwu}. We denote the spaces of such generalised forms by $\O^{p,q}$, where we do not restrict the range of $q$ to be less or equal to that of $p$.  We next define two differentials $\del$ and $\del'$ ($d_L$ and $d_R$ in Hull's notation) which are essentially exterior derivatives but which act only on the first $p$ indices or on the last $q$ indices when applied to a $(p,q)$ form. Hull's field $C$ is a $(2,2)$-form in this notation, while the field strength,
\be
G=\del\del' C, 
\ee
is a $(3,3)$-form. { The Lagrangian for this system is $\tr(G)\cdot C$, where the trace is defined by contracting the first and fourth indices on $G$ using the flat Minkowski metric}. The gauge transformation of $C$ is $\delta C_{2,2}=\del f_{1,2} + \del' f_{2,1}$, where  $f_{2,1}$ is the transpose of $f_{1,2}$. {Note that the properties $\del^2=0= \del'^2$ and $\{\del,\del'\}=0$ ensure that $G$ is invariant under these transformations.} We can consider  {the forms $C$ and $G$} as ultra-local $(p,q)$ forms on $LM$ as in the $(H,B)$ case.

In components a (2,2)-form on loop space $LM$  derived from a (2,2) form on $M$ is given by

\be
C_{mn,pq}(s,t,u,v)[x]=\d(s-t)\d(u-v)\d(s-u) C_{mn,pq}(x(s))\ .
\ee
As a (2,2) form this is

\begin{align}
\label{3.2}
C=&\frac{1}{2.2}\int ds\,dt\,du\,dv\,\big(\d x^{n}(t)\wedge\d x^m(s)\big)\circ \big(\d x^{q}(u)\wedge\d x^{p}(v) \big)\d(s-t)\d(u-v)\d(s-u) C_{mn,pq}(x(s)) \nn\w1
=&\frac{1}{2.2}\int ds\,\big( \d x^n(s)\wedge\d x^m(s)\big)\circ \big(\d x^q(s)\wedge\d x^p(s) \big)C_{mn,pq}(x(s)) \ ,
\end{align}
where $\circ$ indicates that $C$ is a $(2,2)$ form.

We next introduce two vector fields $\dot X$ and $\dot X'$ which generate circle transformations with respect  to the first and second indices respectively.
Contracting $C$ with $i_{\dot X}$ and $i_{\dot X'}$\! \!\footnote{ C.f. left and right contraction in \cite{deMedeiros:2002qpr}.}
we find $\tilde C$ which is given by
\be\label{loopmetric}
\tilde C=i_{\dot X} i_{\dot X'} C=\int ds\, \d x^m(s) \circ\d x^p(s)\dot x^n \dot x^q C_{mn,pq}(x(s))\ ,
\ee
where $\circ$ now indicates the symmetrised tensor product. In components, $\tilde C$ is given by
\be
\tilde C_{mn} (s,t)[x]=\d(s-t)\,\dot x^p(s)\dot x^q(s)\, C_{mp,nq}(x(s))\ .
\ee
This is the result we wanted to obtain: the analogue of the one-form potential on loop space in the $(H,B)$ case is a symmetric two-index tensor in $(G,C)$ theory.
The field strength tensor in loop space can be obtained from the ultra-local loop space version of $G_{3,3}$ in a similar fashion. The formula given in \eq{3.2} can be modified to give
\begin{align}
\label{3.3}
G=&\frac{1}{3!.3!}\int ds\, \d x^p(s)\wedge\d x^n(s)\wedge\d x^{m}(s)\circ \d x^s(s)\wedge\d x^q(s) \wedge\d x^{r}(s)G_{mnp,qrs}(x(s)) \ 
\end{align}
as a local $(3,3)$ form on loop space. We then have
\be
\tilde G=i_{\dot X} i_{\dot X'} G=\frac{1}{2.2}\int ds\, \d x^m(s)\wedge \d x^n(s) \circ\d x^p(s)\wedge \d x^{q}(s) \dot x^r \dot x^s G_{mnr,pqs}(x(s))\ ,
\ee
or in components,
\be
\tilde G_{mn,pq} (s,t,u,v)[x]=\d(s-t)\d(s-u)\d(s-v)\,\dot x^r(s)\dot x^s(s)\, G_{mnr,pqs}(x(s))\ .
\ee

We observe that \re{loopmetric} this differs somewhat from the case of the natural metric on the loop space $LM$ associated with a Riemannian (or Lorentzian) manifold $M$ with metric $g$. In this case the natural metric on $LM$ is given by the  ultra-local expression 
\be 
g_{mn}(s,t)[x]=\d(s-t) g_{mn}(x(s))=\d(s-t)( h_{mn}(x(s))+...)\ ,
\ee
which is similar to the ultra-local two-form field in the $(H,B)$ example above.

\section{Symmetries}

In components, the transformation of $C_{2,2}$ on $M$ is
\be
\d C_{mn,pq}=\del_{[m} f_{n],pq} +( [mn]\leftrightarrow[pq])
\ee
since $f_{2,1}$ is the transpose of $f_{1,2}$. This means that
\be\label{vartildeC} 
\d\tilde C_{mn}(s,t)[x]=\d(s-t) {\dot x}^p{\dot x^q} \left(\del_{[m} f_{p],nq} +( [mp]\leftrightarrow[nq])\right)
\ee

We can express this in terms of an infinitesimal coordinate transformation on $LM$ generated by a local covector field $\xi_m(s)[x]$ given by\footnote{Since we are considering the linearised theory the index on a vector can be lowered by the flat (Minkowski space) metric}.  
\be\label{diff}
\xi_n=\dot x^p(t)\dot x^q(t) f_{p,q n}(x(t)) 
\ee 
We  might then expect  that the transformation of $\tilde C$ can be written as
\be\label{varC2}
\d\tilde C_{mn}(s,t)[x]= {\d \over{\d x^m(s)}} \xi_n(t)[x] + {\d \over{\d x^n(t)}} \xi_m(s)[x] 
\ee
but this is not quite the case as the right-hand side of this equation includes a term involving a derivative of the delta-function $\d(s-t)$ whereas the left-hand side does not. 

We rewrite \re{vartildeC} as 
\bea\nonumber\label{varC}
\d\tilde C_{mn}(s,t)[x]&=&\d(s-t) {\dot x}^p{\dot x^q} \left(\del_{(m} f_{|p|,n)q} -\del_{p} f_{(m,n)q} \right)\\
&=&\d(s-t) \Big({\dot x}^p{\dot x^q} \del_{(m} f_{|p|,n)q} -\dot x^p\frac d{ds} f_{(m,n)p} \Big)
\eea
Using
\be
\frac{\delta x^p(t)}{\delta x^m(s)}=\delta^p_m\delta(s-t)\qquad \Rightarrow \quad \frac{\delta  \dot  x^p(t)}{\delta x^m(s)}=\delta^p_m\delta'(s-t)
\ee
we then evaluate
\be \label{vars}
{\d \over{\d x^m(s)}} \xi_n(t)[x] + {\d \over{\d x^n(t)}} \xi_m(s)[x]\\
=-2\d(s-t) {\dot x}^p{\dot x^q} \left(\del_{(m} f_{|p|,n)q}\right)-2\d'(s-t) {\dot x^p} f_{(m,n)p}
\ee
Combining \re{varC} and \re{vars} we then have
\be
\delta \tilde C_{mn}(s,t)=-{\d \over{\d x^m(s)}} \xi_n(t)[x] - {\d \over{\d x^n(t)}} \xi_m(s)[x]-\dot x^p\frac d{ds} \big(\d(s-t)f_{(m,n)p} \big)\ .
\ee
where $\d\tilde C_{mn}$ is given in \re{varC}. 

As seen from \re{vars} $\d\tilde C_{mn}$ then contains  $\d'$, the derivative of the delta-function (with respect to $s$).
This may be compared to  \cite{Bergshoeff:1991ei} where a derivation of all the constraints of on-shell ten-dimensional supergravity coupled to Yang-Mills is derived using light-like integrability in loop superspace. To cover all couplings it is found that the Yang Mills Kac-Moody algebra being used in this context,
\be
[T_i(s), T_j(t) ] =c_{ij}{}^kT_k(s)\delta (s-t) ~, 
\ee
has to be modified to include a central extension:
\be
[T_i(s), T_j(t) ] =c_{ij}{}^{k}T_k(s)\delta (s-t) +2in\delta' (s-t)~,
\ee
where $n$ is a central charge. So the $\d'$  in \re{vars}  is the corresponding term for our situation.


\section{$(A,B,C)$ theory}

Let us now consider a theory with a non-abelian gauge field $A$, a 2-form potential $B$ and a $C$ field of Hull type. In ordinary spacetime the field strengths are defined by\footnote{Here $g$ is a constant of dimension $-1$.  This  would render the $D=6$ quantum field theory non-renormalisable. However, our discussion is purely classical, and in the string theory context this would not be relevant. }
\bea\nn
&&F=dA +g A^2,\\[1mm]\nn
&&H=dB+Q_3(A)~,\\[1mm]
&&G=\del\del' C+ Q_3\circ H + H_3\circ Q_3~.
\eea
so that the Bianchi identities are
\bea\nn\label{tfs1}
&&\!\!\!\!\!\!\!\!\!\!\!\!\!\!\!\!\!\!\!\!\!\!\!\!\!\!\!\!\!\!\!\!\!\!\!\!DF=0,\\[1mm]\nn
&&\!\!\!\!\!\!\!\!\!\!\!\!\!\!\!\!\!\!\!\!\!\!\!\!\!\!\!\!\!\!\!\!\!\!\!\!dH=P_4(A)\\[1mm]
&&\!\!\!\!\!\!\!\!\!\!\!\!\!\!\!\!\!\!\!\!\!\!\!\!\!\!\!\!\!\!\!\!\!\!\!\!\del\del'G=2 P_4\circ P_4 .
\eea

In loop space we reinterpret $A(s)$ as a gauge field for $LG$, while defining $\tilde H=i_{\dot X}H$, $\tilde B=i_{\dot X} B$ and $\tilde{Q_2}=i_{\dot X}Q_3$ we get
\be
\tilde H= d\tilde B + \tilde{Q}_2(A)\ ,
\ee
and we can interpret $(A(s),\tilde B)$ as the gauge field for $\hat{LG}$ as in the $(B,H)$ section above. For the $C$ field we have two options. On the one hand we can apply $i_{\dot X}$ and $i_{\dot X'}$ to the third row giving the double contracted $G$,  together with $\tilde P_3\circ\tilde P_3$ on the right-hand side or we can interpret $\tilde C_{mn}$ as a metric, construct the corresponding Levi-Civita connection and interpret it as a connection for $LDiff M$. In the second case, forgetting the gauge field $A$ for the moment, we would have an extension of $LDiff M$ to $\hat{L} Diff M$ provided by $\tilde{B}$.
, 

\section{Conclusions}

In this paper we generalise the loop space construction of the $(B,H)$ system to $(C,G)$ where $C$ is a $(2,2)$ and $G$ a $(3,3)$ generalised $(p.q)$ forms. We interpret $\tilde C$ as a linearised metric on loop space and $\tilde G$ as its field strength. The transformations of $\tilde C$ can be written similarly to the usual space time diffeomorphisms up to a term that contains the derivative of a delta function in the circle variables, This has an analogy to the extension of the gauge algebra for a spin one gauge field $A$ coupled to the  $(B,H)$ system. We give arguments why $G$ is invariant under these diffeomorphism transformations. 
In addition we show how the coupling of $A$ can be extended to the $(C,G)$ system.

The introduction of $C$ is originally made in the context of exotic supergravity which is superconformal. In principle we may  lift our results to superspace supergravity simply by reinterpreting indices as superspace indices. There are, however, some issues with this  due to the intepretation of $\tilde C$ as a metric. This is not compatible with Wess Zumino supergravity but  perhaps with  the larger Arnowitt and Nath form of local supersymmetry. However, we have been assuming a Minkowski background for the linearised theory and a supermetric will introduce a scale so that we loose superconformal invariance. One may then reinterpret the construction with an AdS type superspace background which has a scale which we leave for future investigations, We also leave for later the possible route to Wess Zumino supergravity  that entails introducing vielbeins rather than the metric.

We note that the possibility of treating $C$  as giving the norm of two forms was suggested by Hull in \cite{Hull:2000ih}.

\end{document}